\documentclass[summary]{URSIGASS2021}

\title {Study of Low-latitude Ionospheric Scintillation using NavIC}

\author{Sumanjit Chakraborty*\affref{ref1} and Abhirup Datta\affref{ref1},
}

\affiliation{
\aff{ref1}{Department of Astronomy, Astrophysics and Space Engineering, Indian Institute of Technology Indore, Simrol, Indore-453552, Madhya Pradesh, India}
}

\begin{document}

\maketitle

\begin{abstract}

Equatorial ionospheric irregularities have been studied in the past and have produced interesting insights about ionospheric physics and processes. Here, we present the initial results of a long term study of ionosphere near the Equatorial Ionization Anomaly (EIA) using the Navigation with the Indian Constellation (NavIC). We have characterized the ionospheric irregularities in terms of the power spectral density at different dynamical frequencies. The formalism is similar to as suggested by [1-3] using the phase screen modelling of the ionosphere. The observations of the C/N$_o$ (dB-Hz) variation have been taken by utilizing the L5 (1176.45 MHz) signal of NavIC over Indore located near the northern crest of EIA. 
We show some initial results as a proof of concept study from a single day (December 4, 2017) scintillation observations. This is a first of its kind study in this region with NavIC. From the power spectral density analysis, we have demonstrated that NavIC is capable of detecting such irregularities over long periods over this region and has implications in forecasting such events in the future. 

\end{abstract}

\section{Introduction}

Equatorial Plasma Bubbles (EPBs) are the post-sunset plasma instability in the equatorial ionosphere generating irregularities and large scale depletion in the density of the electrons. When the radio waves propagate through these irregularities, they experience scattering and diffraction, causing random fluctuations or scintillations of the Very High Frequency (VHF) signal amplitude, phase, polarization and the propagation direction [4-7]. These scintillation producing irregularities are generally present in and around the F-layer of the ionosphere, where the plasma density is maximum. The primary mechanism responsible for generating the EPBs, with scale sizes from few hundred meters to several kilometers, is the Rayleigh–Taylor (R-T) instability [4-7]. 
Scintillations are observed during the period when solar  activity is at its peak, occurring near the magnetic equator in the post-sunset to midnight sector [8-9]. As these scintillations are demonstration of space weather effects, that affects the performance of space-based navigation systems [10], it becomes crucial to understand the occurrence of scintillation and its after effects. 
Since the early 1950s and following the works by [11-12], characterization of the ionospheric scintillation has been performed by several researchers who developed numerous phase screen theories. Their fundamental approach was the use of wave propagation theory in random media in order to obtain the exiting wave parameters and further solving the problem using theory of Fresnel diffraction that involves radio wave propagation between the ionosphere and the Earth, which finally would replace the ionosphere by an equivalent random phase screen [13]. They considered the ionospheric irregularities as phase objects with the ground pattern produced by the propagating waves through this phase screen derived from the Fresnel diffraction theory [14].

The northern crest of the Equatorial Ionization Anomaly (EIA) and the geomagnetic equator passing the southern tip of the Indian peninsular region, in addition to sharp latitudinal gradients of ionization, makes the Indian longitude sector a highly geosensitive region of investigation for ionospheric research during geomagnetically disturbed periods when the low-latitude ionization is significantly affected as a result of solar eruptions like the Coronal Mass Ejections (CMEs), the Co-rotating Interaction Regions (CIRs) and solar flares [15-16].
Recently, evaluation of the latest versions of the existing global empirical ionospheric models, the International Reference Ionosphere 2016, the NeQuick2 and the IRI-extended to Plasmasphere, are shown in details to be incapable of predicting the geomagnetically disturbed conditions over the Indian longitude sector [17], thus suggesting the need for investigating the existing knowledge of the phase screen model for proper characterization of the low-latitude ionosphere over the Indian sector. 
In this paper for the first time, we show initial results by performing the power spectral density analysis and demonstrate that NavIC is capable of detecting such irregularities over long periods over this region, which would have implications in forecasting such events in the future.

\section{Data}

Data has been analysed using a NavIC receiver, provided by the SAC-ISRO, capable of receiving L5 (1176.45 MHz) signal, operational since 2017, in the Department of Astronomy, Astrophysics and Space Engineering of IIT Indore.

\section{Results and Discussions}

Due to the passage of a High Speed Solar Wind (HSSW) stream around the Earth, arising from a positive polarity CH on the Sun, a G1 (K$_p$= 5, minor) level geomagnetic storm was observed on December 4, 2017 according to the NOAA. The Dst index showed dip during December 4, 2017 with a value -45 nT at 22:00 UT.

\begin{figure}[ht]
\centering
\includegraphics[width=3.3in,height=3.25in]{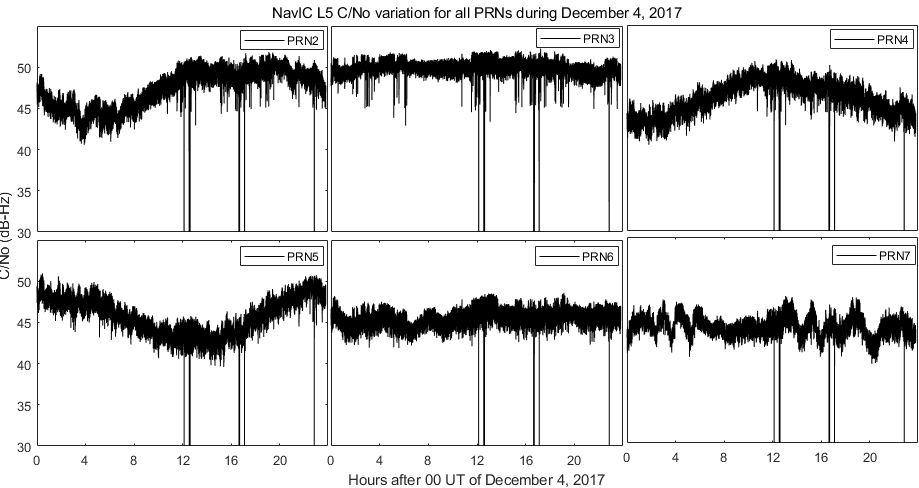}
\caption{The C/N$_o$ (dB-Hz) variation during the entire day of December 04, 2017, as observed by L5 signal of NavIC satellite PRNs 2-7.The C/N$_o$ values that had dropped to zero are designated here as drops to 30 dB-Hz.} 
\label{sc1}
\end{figure}

Figure \ref{sc1} shows the C/N$_o$ variation (dB-Hz) for the entire day of December 4, 2017, as observed by the L5 signal of NavIC. Drops in the C/N$_o$ has been observed at multiple time stamps throughout the day by all the PRNs (consisting of a set of geostationary and geosynchronous satellites). It is to noted that the C/N$_o$ values had dropped to zero in these time-stamps and that they are designated in the figure as drop-downs to 30 dB-Hz. 

\begin{figure}[ht]
\centering
\includegraphics[width=3.3in,height=3in]{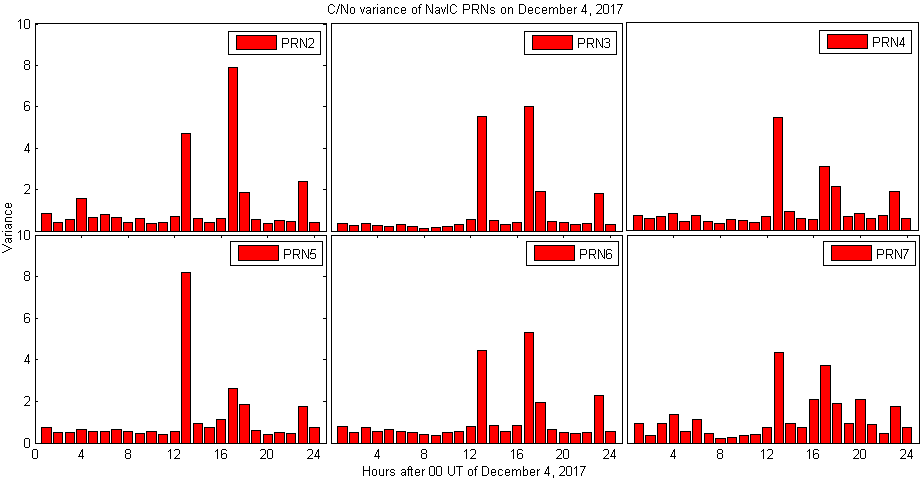}
\caption{The hourly binned variance plots of C/N$_o$ for all PRNs of NavIC on December 4, 2017. Most significant rises in variance values as observed from PRN 2 (16-17 UT bin) and PRN 5 (12-13 UT bin).} 
\label{sc2}
\end{figure}

However, for the identification of significant C/N$_o$ drops, Figure \ref{sc2} shows the hourly binned variance plots of all the PRNs of NavIC during the entire day of December 4, 2017. The bin of 16-17 UT for the PRN 2 and that of 12-13 UT for PRN 5 show the most significant rise and hence maximum variation among all the bins of the day.

The PSD is given by [1-3]: 
\begin{equation}
S = \frac{\alpha}{(f_0^2 + f^2)^\frac{\beta}{2}}
\label{sca}
\end{equation}
where $\alpha$ is the spectral strength, f is the dynamical frequency and $\beta$ is the spectral slope. 

For $f >> f_0$, equation \ref{sca} becomes simplified to:
\begin{equation}
S = \alpha f^{-\beta}
\label{scb}
\end{equation}
 
\begin{figure}[ht]
\centering
\includegraphics[width=3.3in,height=2.5in]{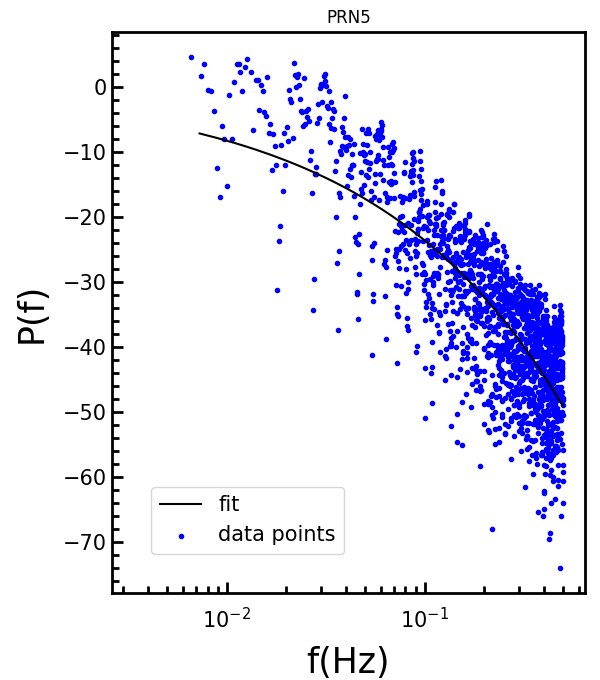}
\caption{The PSD variation with the least square fit (black solid line) and the smoothed data (red solid line) corresponding to the C/N$_o$ variation designated in horizontal bars in \ref{sc2} as observed from PRN 5 during December 4, 2017.}
\label{sc3}
\end{figure}

Figure \ref{sc3} depicts the PSD variation from one of the PRNs (PRN 5) which shows a power-law variation in the PSD and thus is taken into consideration for calculation of the $\beta$ which is found out to be 0.459$\pm$0.039, showing occurrence of weak scintillation that is detected by NavIC.   
 
\section{Conclusions}

The ionosphere over the Indian longitude sector is highly dynamic and geosensitive as a result of presence of the northern crest of EIA. 
Previous studies show development of a power law phase screen theory and models for scintillation based on weak and strong conditions. However, nature of the phase screen associated with  scattering, near the EIA of the geosensitive region of India, with observations utilizing the NavIC satellites' signal (L5:1176.45 MHz), has not been studied in the literature. 
To address the problem, this study was performed during December 4, 2017 that caused fluctuations in the NavIC C/N$_o$ variation. The spectral slope from the PSD was obtained to be 0.459$\pm$0.039. 
This paper for the first time, demonstrated the capability of NavIC for detecting such irregularities from the PSD analysis. This single day proof of concept study will be followed up with a detailed study of scintillation analysis.  

\section{Acknowledgements}

The authors acknowledge Space Applications Centre, Indian Space Research Organization for providing the NavIC receivers to Department of Astronomy, Astrophysics and Space Engineering, IIT Indore under the project NGP-17. Further acknowledgements go to World Data Center for Geomagnetism, Kyoto for the Dst index freely available at http://wdc.kugi.kyoto-u.ac.jp/kp/index.html.


\end{document}